\documentclass[prb,aps,epsf,twocolumn,superscriptaddress,showpacs,10pt,]{revtex4-2}
\usepackage{graphicx,amsfonts,times,bm,amsmath,verbatim,color,array}
\usepackage{xcolor}
\usepackage{algorithmic}

\usepackage[colorlinks,urlcolor=blue,citecolor=blue,linkcolor=blue]{hyperref}
\usepackage[capitalize]{cleveref}

\usepackage{multirow}
\usepackage[all]{hypcap} 
\usepackage{xspace}
\usepackage{amssymb}
\usepackage{appendix}
\usepackage{pifont}
\usepackage{booktabs}
\usepackage{etoolbox} 
\usepackage{lipsum} 
\usepackage[capitalize]{cleveref}

\usepackage{amsmath}   
\usepackage{mathrsfs}  

\AtBeginDocument{%
  \heavyrulewidth=.08em
  \lightrulewidth=.05em
  \cmidrulewidth=.03em
  \belowrulesep=.65ex
  \belowbottomsep=0pt
  \aboverulesep=.4ex
  \abovetopsep=0pt
  \cmidrulesep=\doublerulesep
  \cmidrulekern=.5em
  \defaultaddspace=.5em
}

\begin{document}

\title{Magnetic field Controlled Anderson Delocalization in a Spinful Non-Hermitian Chain}

\author{Moirangthem Sanahal}
\email{sanahal1523o@gmail.com}
\affiliation{Department of Physics, National Institute of Technology Silchar, Assam 788010, India}

\author{Subhasis Panda}
\email{subhasis@phy.nits.ac.in}
\affiliation{Department of Physics, National Institute of Technology Silchar, Assam 788010, India}

\author{Snehasish Nandy}
\email{snehasish@phy.nits.ac.in}
\affiliation{Department of Physics, National Institute of Technology Silchar, Assam 788010, India}

\begin{abstract}
Anderson localization (AL) and the non-Hermitian skin effect (NHSE) represent two paradigmatic localization phenomena driven, respectively, by disorder and non-Hermiticity. In one-dimensional (1D) non-Hermitian systems, these factors are known to compete and provide a smooth crossover between AL and NHSE upon parameter tuning. Here, we show that this interplay is fundamentally enriched in spinful systems, where an external magnetic field acts as an additional degree to manipulate the localization behavior. By investigating a disordered 1D spinful non-Hermitian chain, we demonstrate that under appropriately correlated disorder configurations across spin sectors, the magnetic field enhances the AL $\rightarrow$ NHSE crossover. Interestingly, this facilitates the Anderson delocalization transition even in strongly disordered systems where states would otherwise be Anderson localized. By analyzing the inverse participation ratio and the mean center of mass, we map the resulting triple interplay between disorder, non-Hermiticity, and the magnetic field strength, identifying regimes of Anderson localization and skin accumulation. We further reveal that this magnetic field driven delocalization phenomenon originates from an effective suppression of disorder strength via Zeeman-induced inter-chain coupling across the spin sectors. 
\end{abstract}

\maketitle

\section{Introduction}

Non-Hermitian physics has emerged as an effective theoretical framework for describing realistic physical systems that are intrinsically open, dissipative, or driven and thus lie beyond the paradigm of idealized Hermitian models. Originally motivated from the observation that $PT$ (parity-time reversal) symmetric non-Hermitian (NH) Hamiltonians can have real eigenspectra~\cite{bender1998}, non-Hermitian physics has since garnered widespread interest across diverse platforms, ranging from optical~\cite{NHSE_via_gainloss_in_opticlly_coupled_cavity_array,Observing_PTsym_in_optics,EPs_enhance_sensing_in_optical_microcavity,topological_funelling,optical_force_enhancement,twist_induced_NHSE_in_waveguides}, mechanical~\cite{Observation_of_NHSE-in_mechanical_metamaterial}, electrical~\cite{Gen_BBC_in_NH_topoelectric_circuit,nonAbelian_gauge_fields_in_circuit,HOSE_in_NH_topoelectric_circuit,observation_of_SDBE_in_NH_electric,ReciprocalSE_and_its_realization,NHSE_in_a_NH_electrical_circuit}, photonic~\cite{manipulate_NHSE_by_ORR,PT_sym_photonic_lattice,PT_sym_microring_lasers,PT_sym_whispering_gallery_microcavities,NH_photonics_based_on_PTsym,nonAbelian_effects_in_dissipative_photonics,GDSE_in_reciprocal_photonic}, ultra-cold~\cite{NHSE_in_UCatoms,engineering_NHSE_in_UCatoms,DYnamic_signs_of_NHSE_in_UCatoms,theoretical_prediction_of_NHSE_in_UCatom,NHSE_in_periodically_driven_dissipative_UCatom,2D_NHSE_in_UCfermigas} to open quantum systems~\cite{Phy_of_open_QM_systems,NHSE_in_open_QM_systems,Liouvillian_SE_in_exactly_solvable_model}. In the field of NH topology, non-Hermiticity is studied vastly through the lens of condensed matter physics, uncovering unique phenomena that have no Hermitian counterparts~\cite{symmetry&topology,topPhasesofNHsystem,NH_topological_phenomena:A_review,principles&prospects,NH_topology_ib_H_matter}. One such is the non-Hermitian skin effect (NHSE), marked by an extensive boundary localization of bulk states under open boundary conditions (OBC) and originating from a point gap topology in the complex energy eigenspectrum~\cite{anomalousedgestates,topologicalorigin,correspondence,areview,aperspective}. 

On the other hand, Anderson localization (AL), as an old yet profoundly influential phenomenon, remains an important branch in Hermitian condensed matter physics~\cite{PW_Anderson}. Unlike the boundary-localization driven by non-Hermiticity in NH systems, AL represents localization throughout the bulk driven by disorder in Hermitian systems. In recent years, this paradigm has been revisited in the broader context of disordered non-Hermitian systems, where non-Hermiticity fundamentally alters localization behavior~\cite{localisationtraninNHQM,interplay_of_NHSE_and_AL,AL_transitions_in_disordered_NH,AL_in_NH_AAH_model,Metal_Ins_phase_transition_in_NH_AAH_model,NHSE_&_WN_in_disordered_NH_systems}. A seminal example is the work by Hatano and Nelson, which revealed that non-reciprocal hopping can induce an Anderson delocalization transition in a disordered one-dimensional (1D) NH chain, accompanied by the emergence of NHSE~\cite{localisationtraninNHQM}. This behavior challenged the Hermitian understanding that 1D systems are always Anderson localized, even for arbitrarily weak disorder~\cite{scaling_theory_of_localization}. Disorder and non-Hermiticity do not act independently in non-Hermitian systems; rather, they compete. At weak disorder, the NHSE dominates, whereas increasing disorder strength progressively suppresses the skin modes, eventually giving way to AL. 

Beyond this direct competition between disorder and non-Hermiticity, it has been recently shown that the AL-NHSE localization transition can be controlled via an alternative route. Anderson localization in a strongly disordered NH chain can be suppressed by coupling it to a Hermitian chain with appropriately correlated disorder~\cite{AD_in_strongly_coupled}. This inter-chain coupling induced revival of the NHSE introduces an additional degree of freedom, thereby opening a richer three-way interplay between \textit{disorder}, \textit{non-Hermiticity}, and \textit{inter-chain coupling}. Motivated by this result, a pertinent question arises: can a similar delocalization transition be realized by introducing coupling between two disordered NH chains? Such NH-NH chain coupling underlies several NH phenomena, including the Critical NHSE~\cite{critical_NHSE_Nature,scaling_rule_for_cNHSE}, and the unidirectional accumulation of bilocalized skin modes in spinful systems~\cite{sanahal1}. Moreover, in spinful platforms, this inter-chain (spin) coupling can be physically realized through an external in-plane magnetic field via Zeeman interaction, providing a simple, physically accessible and experimentally relevant route to engineer NH-NH chain coupling. Thus, we ask: can an external magnetic field drive Anderson delocalization even in strongly disordered NH systems? From this perspective, we investigate the resulting triple interplay between \textit{disorder}, \textit{non-Hermiticity}, and \textit{magnetic field} in 1D spinful NH systems with correlated disorder.

In this work, we provide affirmative answers to the above questions and demonstrate the notions of localization using a minimal spinful extension of the Hatano-Nelson model. We first illustrate a smooth crossover between AL and NHSE, arising from the competition between \textit{disorder} and \textit{non-Hermiticity}. We then demonstrate that, in the spinful setting, the introduction of an external in-plane magnetic field significantly enhances the AL $\rightarrow$ NHSE transition: driving Anderson delocalization even in the presence of strong disorder. We uncover that this magnetic field induced resurgence of the NHSE occurs because of an effective suppression of disorder strength by the magnetic field. Finally, the resulting triple interplay between disorder, non-Hermiticity, and magnetic field is quantitatively characterized using the `inverse participation ratio' (IPR) and the `mean center of mass' (mcom). We find that stronger \textit{disorder} require a stronger magnetic field to induce Anderson delocalization, while increasing \textit{non-Hermiticity} reduces the field strength required to trigger the same.

The rest of the paper is organized as follows: we start with a description of the model Hamiltonian in Sec.~\ref{sec: Model}, under both spinless and spinful frameworks. The resulting observations under both the cases are highlighted in Sec.~\ref{sec: Results}, which includes the competition between disorder and non-Hermiticity in the spinless case, and the effect and role of magnetic field in altering them, in the spinful extension. Sec.~\ref{sec: mechanism underlying magnetic field driven Anderson delocalization} demonstrates the mechanism behind the magnetic field driven Anderson delocalization phenomenon observed in the spinful case. Sec.~\ref{sec: Conclusion} finally concludes by summarizing our results and
providing an outlook.

\section{Model Hamiltonian \label{sec: Model}}
In this section, we employ minimal models in both spinless and spinful frameworks to investigate the interplay between disorder and non-Hermiticity in one dimension. We begin with the 1D Hatano-Nelson model as the prototypical spinless case, then extend it to a spinful formulation by incorporating synthetic spin-dependent gauge fields. These additional spin degrees of freedom allow us to introduce an external magnetic field as a tunable parameter, providing us control over the system's localization behavior.

\begin{figure}[b]
    \centering
    \vspace{0.18cm}
    \includegraphics[width=1\linewidth]{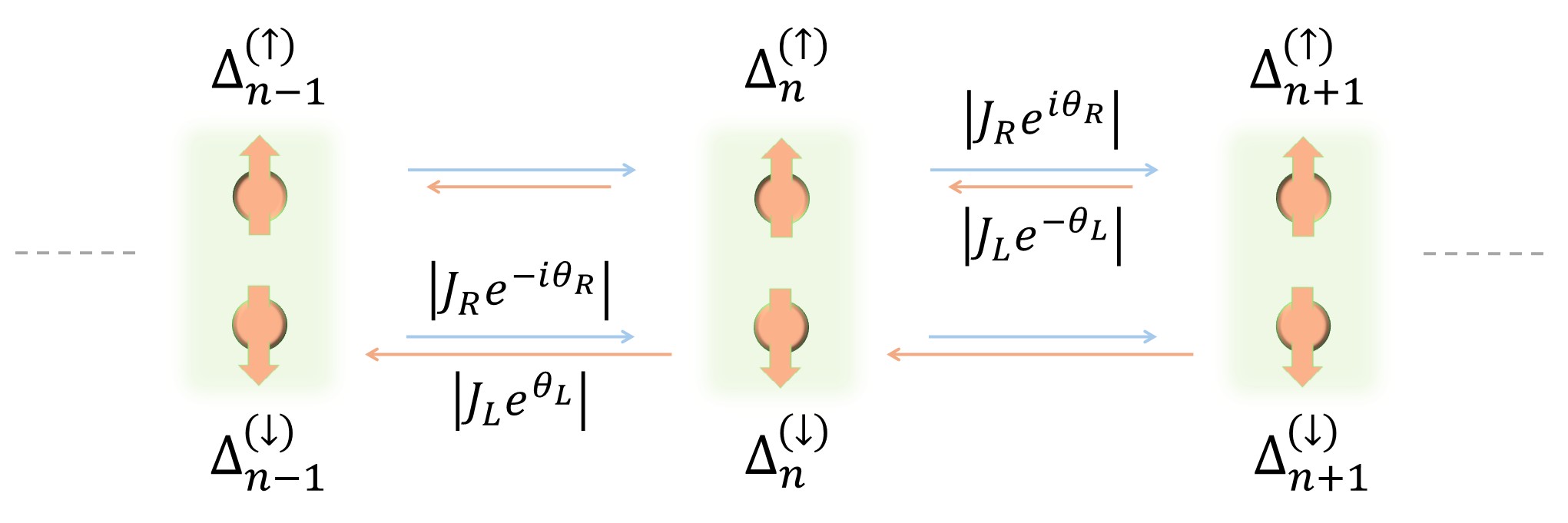}
    \caption{Schematic of a disordered spinful 1D Hatano-Nelson model subjected to gauge type $\mu,\nu = z$ such that $\theta_L$ is chosen to be imaginary.}
    \label{fig: Model Schematic}
\end{figure}

\subsection{Spinless chain}
The Hamiltonian of the 1D Hatano-Nelson model in the presence of onsite disorder has the form~\cite{localisationtraninNHQM}
\begin{equation}
\begin{aligned}
    \mathcal{H} = \sum_{n=1}^{N-1} \left( t_L c_{n}^\dagger c_{n+1} + t_R c_{n+1}^\dagger c_{n} \right) + \sum_{n=1}^{N} \Delta_n c_n^{\dagger}c_n,
\end{aligned}
\label{eqn:spinless HN model}
\end{equation}
where, $c_n^{\dagger}$ and $c_n$ are the creation and annihilation operators at site $n$. The parameters $t_L,t_R$ represent leftward, rightward nearest-neighbor hopping respectively, whose imbalance encodes the system's non-Hermiticity, and $\Delta_n$ denotes a site-dependent random onsite potential at $n^{th}$ site. The disorder in the system arises through the spatial fluctuation of the potential across the lattice sites, with their values $\{ \Delta_n\}$ sampled uniformly from the interval $\left[ -W/2,W/2 \right]$. The extent of fluctuation, captured by the width `$W$' of the interval, quantifies the disorder strength. 

\subsection{Spinful chain}
Extending the model to a spinful setting, we incorporate the internal spin degrees of freedom by employing spin-dependent gauge fields. With onsite disorder present, the generic Hamiltonian reads~\cite{sanahal1,synthetic_non_abelian,synthesis&observation_of_nonAbelian_gauge_fields_in_real_space,nonabelian_lattice_gaugefields_in_phtonic...}
\begin{align}
    \label{eqn: spinful HN model}
    \mathcal{H} = &\sum_{n=1}^{N-1} (J_Le^{iA_{L}} c_{n}^\dagger c_{n+1} + J_R e^{iA_{R}} c_{n+1}^\dagger c_{n})  \ + \\
    & \sum_{n=1}^{N} (\Delta_n^{(\uparrow)} c_{n,\uparrow}^{\dagger}c_{n,\uparrow} + \Delta_n^{(\downarrow)} c_{n,\downarrow}^{\dagger}c_{n,\downarrow}) + \sum_{n=1}^{N} \bm{B}\cdot \bm{\sigma} \, c_n^{\dagger}c_n \nonumber .
\end{align}

The matrix-valued couplings $J_Le^{iA_L}$ and $J_Re^{iA_R}$ describe the spin dependent hopping of a spin $1/2$ particle on the chain, where $A_{L,R}$ are arbitrary matrix-valued gauge potentials of the form $\theta_{L,R}\sigma_{L,R}$. Here, $\theta_{L,R}$ and $\sigma_{L,R}$ denote the gauge fluxes and Pauli matrices, associated with leftward, rightward hopping in the lattice~\cite{nonabelian_lattice_gaugefields_in_phtonic...,synthesis&observation_of_nonAbelian_gauge_fields_in_real_space}. The terms $\Delta_n^{(\uparrow)}$ and $\Delta_n^{(\downarrow)}$ $(\in \mathbb{R})$ represent random onsite potentials at site $n$ in the two spin sectors, and $\bm{B}$ denotes an external magnetic field. This field enters the system exclusively through the Zeeman term $\bm{B}\cdot \bm{\sigma}$; coupling directly to the spins as orbital effects are absent in one dimension.

The inclusion of these spin degrees effectively maps the model onto a two-chain system. We consider Abelian gauge fields: $\sigma_L,\sigma_R = \sigma_z$ for simplicity, as they decouple the two spin sectors (chains) in the absence of the magnetic field and facilitate independent control of each. Also, to isolate the role of gauge fields in the spinful scenario, we set $J_L=J_R=J$ and embed the non-Hermiticity solely in the gauge fluxes $\theta_{L,R}$. Fig.~\ref{fig: Model Schematic} shows a schematic of the model under this gauge configuration.

\section{Results \label{sec: Results}}
While the spinless framework sufficiently captures the fundamental competition between disorder and non-Hermiticity, the inclusion of internal spin degrees of freedom and an external magnetic field significantly enriches the physics. In this section, we systematically present our numerical results. We begin by revisiting the interplay between disorder and non-Hermiticity within the spinless framework. We then generalize the analysis to the spinful case, where we demonstrate that an external magnetic field fundamentally alters the localization properties; inducing Anderson delocalization in strongly disordered regimes, despite the system being disorder-dominant. Finally, we elucidate the resulting three-way interplay between disorder, non-Hermiticity, and the magnetic field.

\subsection{Competition between disorder and non-Hermiticity \label{sub sec: Competition between non-Hermiticity and disorder}}
In Hermitian regime, it is well established that 1D disordered systems are always Anderson localized, with even infinitesimal disorder triggering the effect~\cite{scaling_theory_of_localization}. This conventional behavior fundamentally alters in NH regimes, where asymmetric couplings modify the localization behavior~\cite{localisationtraninNHQM}. In particular, disorder and non-Hermiticity directly compete with each other; depending on their relative strength, the system may host Anderson localized states, non-Hermitian skin states, or a coexistence of both. Fig.~\ref{fig:Representative NHSE, AL}a illustrates this competition via a parameter plot, identifying distinct localization regimes of the spinless Hatano-Nelson model (Eqn.~\ref{eqn:spinless HN model}) as a function of disorder and non-Hermiticity strength. 

\begin{figure}[t]
    \centering
    \includegraphics[width=1\linewidth]{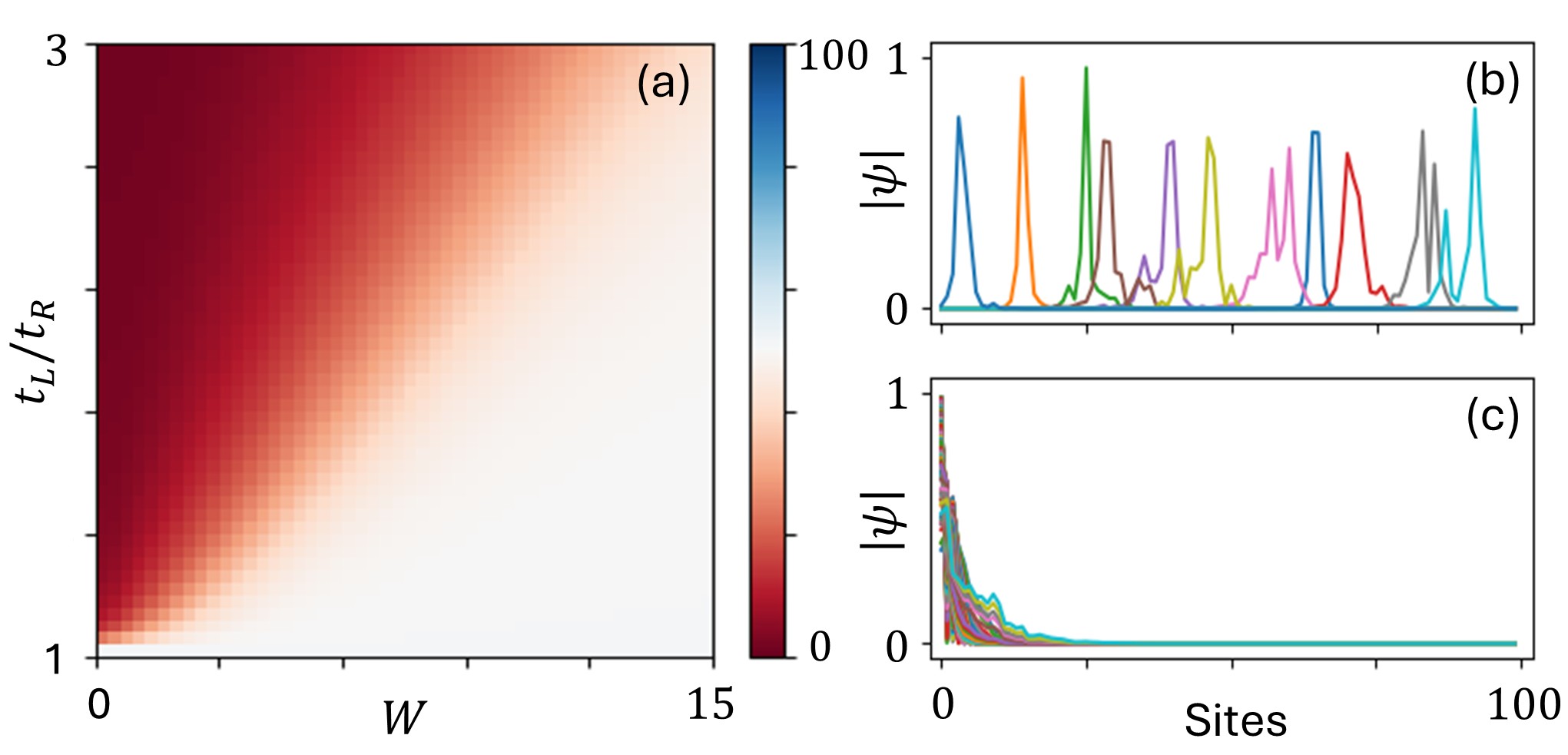}
    \caption{\textbf{(a)} Parameter plot illustrating localization regimes, obtained using $\langle \overline{\text{mcom}} \rangle$ for the spinless model with $t_R=1$ and $N=100$. The red region represents non-Hermiticity dominated region with left-localized skin states and white represents disorder dominated (or clean) regime hosting Anderson (or extended) states. \textbf{(b)} Representative Anderson states and \textbf{(c)} skin states in the disorder dominated and NH dominated region, respectively, obtained from one disorder realization. The result in \textbf{(a)} is averaged over 1000 disorder realizations. The `$\text{mcom}$' is discussed in Sec.~\ref{sub sec Triple interplay of disorder, non-Hermiticity, and Magnetic field} in detail. }
    \label{fig:Representative NHSE, AL}
\end{figure}

In the Hermitian limit ($t_L=t_R$), eigenstates exhibit Anderson localization under finite disorder (extended in the clean limit), as expected and supported by the white horizon in the parameter plot. Upon entering the NH regime ($t_L\neq t_R$), the localization behavior becomes sensitive to the relative strengths of disorder and non-Hermiticity. In regimes relatively dominated by disorder, the states are Anderson-localized, whereas in the non-Hermiticity dominated ones, states are skin-localized. We refer to these regimes as the strong disorder and weak disorder regimes, dictated by the white and red regions, respectively, in Fig.~\ref{fig:Representative NHSE, AL}a. Importantly, these regimes are not separated by a sharp phase boundary; instead, they are connected by a smooth crossover upon parameter tuning. The intermediate regimes exhibit coexistence of both Anderson and skin states, reflecting the continuous competition between the disorder-induced localization and non-Hermitian pumping. Nevertheless, one may still delineate these regimes through a scaling analysis of the relevant localization lengths, as discussed in Ref.~\cite{NH_multipole_SE_challenges_loc}.
Representative eigenstate profiles corresponding to the disorder dominated and non-Hermiticity dominated regimes are illustrated in Figs.~\ref{fig:Representative NHSE, AL}b and~\ref{fig:Representative NHSE, AL}c, respectively. 

As demonstrated by Fig.~\ref{fig:Representative NHSE, AL}a, strong disorder requires relatively stronger non-Hermiticity to drive AL$\rightarrow$NHSE crossover. Conversely, increasing non-Hermiticity necessitates stronger disorder to restore AL. This smooth and continuous competition establishes a clear comparative baseline for understanding how additional spin degrees of freedom and the external magnetic field would modify the localization behavior, which is the focus in the following subsection.

\begin{figure}[b]
    \centering
    \includegraphics[width=1\linewidth]{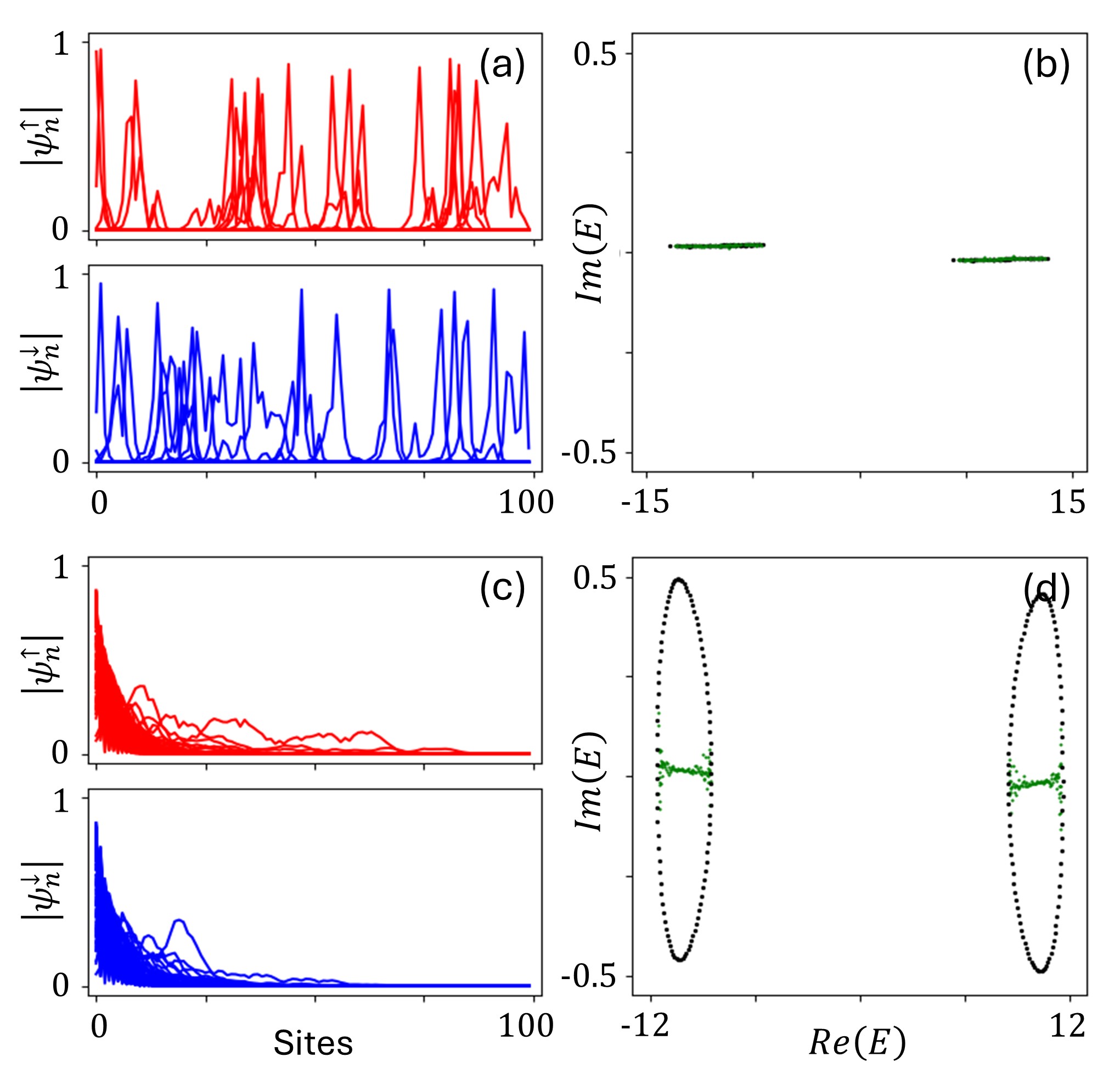}
    \caption{\textbf{(a-b)} Eigenstates: $|\psi_n^{\uparrow}|, |\psi_n^{\downarrow}|$, and corresponding eigenspectrum of the system under external magnetic field ($B\hat{i}$) in the presence of symmetrically correlated disorder. The eigenstates retain Anderson localization even under the magnetic field influence. Only a few eigenstates have been plotted for clarity. \textbf{(c-d)} Similar plots for the case of anti-symmetrically correlated disorder. All the eigenstates exhibit NHSE following Anderson delocalization. The eigenspectrum \textbf{(d)} exhibits point gap with PBC (black) eigenenergies enclosing the OBC (green) eigenenergies. Parameters chosen include $(\theta_L,\theta_R) = (i/5, 1)$, $W/J=5$, $B/J=10$, and $N=100$. All of the above results are numerically obtained from one representative disorder realization. }
    \label{fig:summetric, anti-symmetric}
\end{figure}

\subsection{Magnetic field driven Anderson delocalization \label{sub sec: Magnetic field driven Anderson delocalization}}

We now turn towards the spinful extension of the model, where we elevate the internal spin degrees using $e^{i\theta_L\sigma_z}$, $e^{i\theta_R\sigma_z}$ gauge fields, without any loss of generality. To engineer non-reciprocity in the chain, we choose $\theta_L\in \mathbb{C}$ as purely imaginary. These ingredients render the model equivalent to two decoupled chains with biased hopping amplitudes, as indicated in Fig.~\ref{fig: Model Schematic}. An external magnetic field applied along the in-plane ($B\hat{i}$ say) directly couples to the spins at each site via the Zeeman term ($B\sigma_x$) to give rise to onsite spin–spin interactions (inter-chain couplings) through its off-diagonal elements.

In the absence of the magnetic field, the two spin sectors remain independent and behave equivalently to spinless chains. Under strong disorder, both sectors exhibit AL, and increasing non-Hermiticity drives the familiar AL$\rightarrow$NHSE crossover, following an Anderson delocalization. Interestingly, our central result unveils that this Anderson delocalization crossover can alternatively be achieved by applying an external in-plane magnetic field. The phenomenon, however, requires appropriately correlated disorders in the two spin chains. Disorders of the type $\Delta_n^{(\uparrow)} = \Delta_n^{(\downarrow)}$ are said to be symmetrically correlated and $\Delta_n^{(\uparrow)} = -\Delta_n^{(\downarrow)}$; anti-symmetrically correlated. 

The magnetic field induced effects observed under different disorder configurations are shown in Fig.~\ref{fig:summetric, anti-symmetric} which illustrates the resulting eigenstates $|\psi_n^{\uparrow}|$, $|\psi_n^{\downarrow}|$ and the corresponding eigenspectra of the system under strong disorder. For uncorrelated and symmetrically correlated disorder configurations, the coupled system remains Anderson localized, which is also reflected by the absence of point gap in the corresponding eigenspectrum. In contrast, under anti-symmetrically correlated disorder, the states undergo Anderson delocalization and accumulate at a system boundary, unveiling a re-emergent NHSE. This transition is corroborated by the formation of point gaps in the complex eigenspectrum, where the PBC (periodic boundary condition) spectrum encircles the OBC (open boundary condition) eigenenergies~\cite{topologicalorigin,topPhasesofNHsystem,correspondence}.   

This result establishes the magnetic field as an additional tuning parameter capable of controlling localization behavior in disordered NH systems without even altering their intrinsic disorder or non-Hermiticity, most appreciably driving Anderson delocalization when strongly disordered. The driving mechanism behind this delocalization transition is discussed in Sec.~\ref{sec: mechanism underlying magnetic field driven Anderson delocalization}.

\subsection{Triple interplay of disorder, non-Hermiticity and \\
Magnetic field \label{sub sec Triple interplay of disorder, non-Hermiticity, and Magnetic field}}
Next, we quantitatively characterize the competing effects of \textit{disorder}, \textit{non-Hermiticity}, and \textit{magnetic field} by numerically computing the quantifiers: \textit{inverse participation ratio} (IPR), and the \textit{mean center of mass} (mcom). The IPR of a $m^{\rm th}$ eigenstate is defined as~\cite{midya,molignini,AD_in_strongly_coupled}
\begin{equation} 
    \text{IPR}_m = \frac{\sum_{n=1}^{N} |\psi_m(n)|^4 }{ \left( \sum_{n=1}^{N}  |\psi_m(n)|^2  \right)^2},
\end{equation} 
with $n$ indexing the discrete site indices. It measures the degree of localization of the state: a value approaching 1 indicates maximal localization, while lesser values of the order $\mathcal{O}(1/M)$ indicate relatively weak localized states spread over $M\ll N$ sites, giving IPR $\simeq 1/M$.  In a normalized eigenbasis, the unity is reached exclusively by a perfectly single site-localized (Dirac delta) eigenstate, and IPR $=0$ is reached by a fully extended state in the thermodynamic limit ($N\rightarrow\infty$). Thus, the IPR depends both on the intrinsic nature of the state and the system size.

The IPR, however, cannot distinguish skin localization from bulk localization. To resolve this, we employ the mcom~\cite{molignini,AD_in_strongly_coupled}:
\begin{equation}
    \text{mcom}_m = \frac{\sum_{n=1}^{N} \left( n\ |\psi_m(n)|^2 \right)}{\sum_{n=1}^{N} |\psi_m(n)|^2}.
\end{equation} 

The mcom measures the probability-weighted \textit{average spatial index} of the eigenstate. Values close to $1$ ($N$) indicate strong localization near the left (right) boundary, signaling NHSE whereas values near $N/2$ suggest either an extended state or a state centrally-localized in the bulk. The latter two scenarios can be relatively distinguished by their IPR values, as extended states typically exhibit negligible $\text{IPR}$. For multiple bulk-localized states that are uniformly scattered across the system (AL), their average mcom also tends to $N/2$. In essence, the $\text{IPR}$ excels in detecting AL, while the $\text{mcom}$ better accounts the directional bias associated with the NHSE.

Thus, a combined analysis of the IPR and mcom --- averaged over all eigenstates (i.e. $\overline{\text{IPR}}$ and $\overline{\text{mcom}}$) --- serves as a direct and collective diagnostic to distinguish NHSE from AL in the system as a whole. Further averaging them over multiple disorder realizations for precision, yields
\begin{equation}
    \left<\overline{\text{IPR}}\right> = \left< \frac{1}{2N} \sum_{m=1}^{2N} \text{IPR}_m \right>,
\end{equation} 
\begin{equation}
    \left<\overline{\text{mcom}}\right> = \frac{\sum_{n=1}^N n\left<P(n)\right>}{\sum_{n=1}^N \left<P(n)\right>}.
\end{equation} 

Here,
\begin{equation*}
    P(n) = \frac{1}{2N} \sum_{m=1}^{2N} \left(|\psi_m(n)|^2 \right)
\end{equation*} 
denotes the averaged spatial probability at $n^{{\rm th}}$ site, accounting for the full set of $2N$ eigenstates, and $\left<\#\right> = \sum_{j=1}^{M}(\#_j)/ M$ represents the disorder average, with $M$ as the ensemble size. We remark that a measure of the winding number could also serve as a probe for localization, but it is qualitatively binary in nature. It indicates either an absence or a presence of NHSE. It does not quantify the accumulation strength and thus cannot highlight the regions of AL-NHSE coexistence. This distinction can be made using the mcom.

\begin{widetext}

\begin{figure}[t]
    \centering
    \includegraphics[width=0.9\columnwidth]{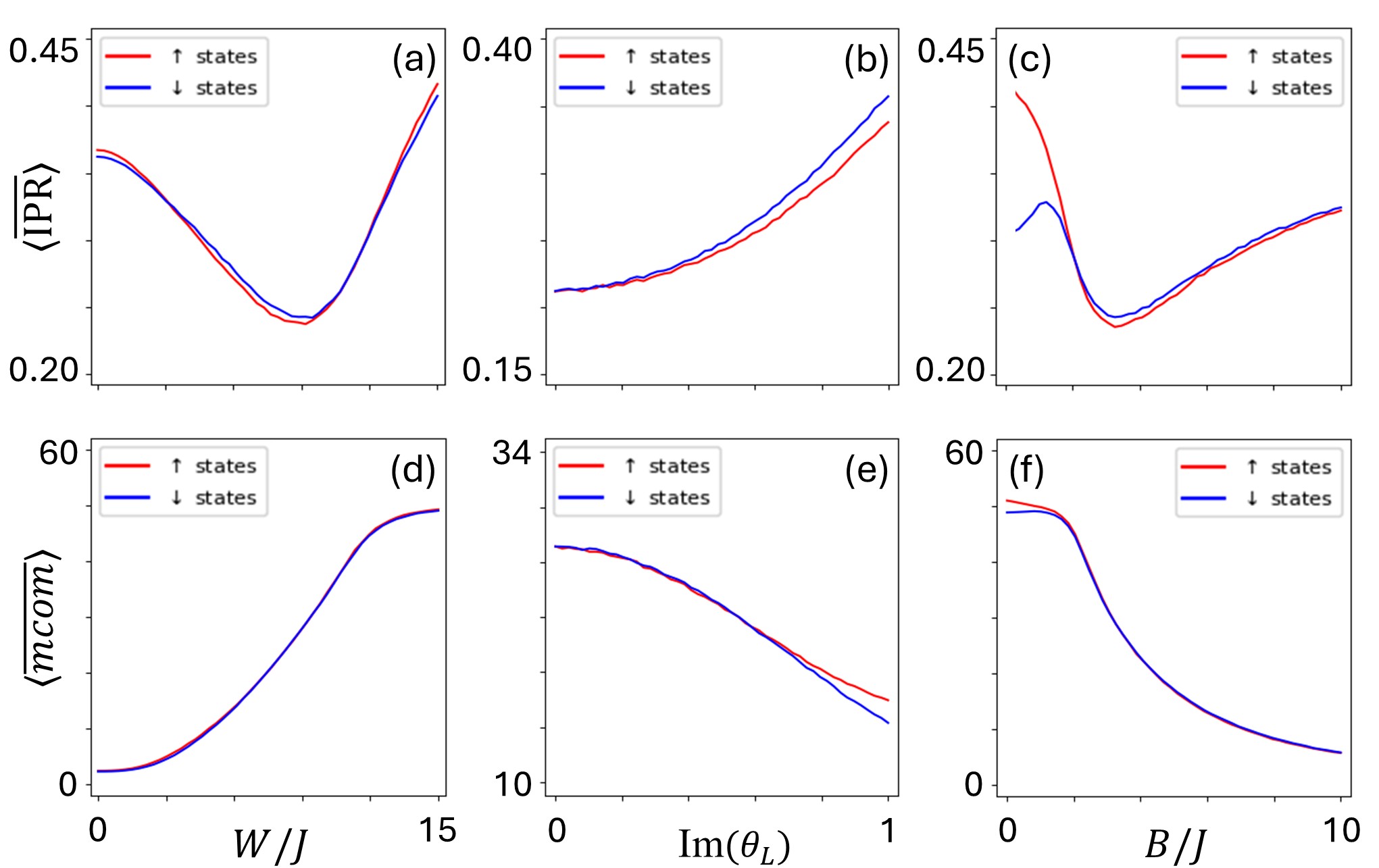}
    \caption{$\langle \overline{\text{IPR}} \rangle$  and $\langle \overline{\text{mcom}} \rangle$ plots. \textbf{(a-c)} Upper panel represents the $\langle \overline{\text{IPR}} \rangle$ and \textbf{(d-f)} lower panel represents the $\langle \overline{\text{mcom}} \rangle$ for both up- and down-spin sectors, under anti-symmetrically correlated disorder as a function of \textbf{(a,d)}  disorder with $\theta_L=i/2$, $B/J=4$, \textbf{(b,e)} non-Hermiticity with $B/J=4$, $W/J=8$, and \textbf{(c,f)} magnetic field with $\theta_L=i/2$, $W/J=8$. The red (blue) curves indicate up(down)-spin sectors. The results are averaged over an ensemble of 1000 disorder configurations. Common parameters include $\theta_R=1$ and $N=100$.}
    \label{fig:IPR,mcom}
\end{figure}

\end{widetext}


 Figure~\ref{fig:IPR,mcom} presents the $\langle \overline{\text{IPR}} \rangle$ and $\langle \overline{\text{mcom}} \rangle$ for both spin sectors under anti-symmetrically correlated disorder. Panels (a) and (d) together track the competition between disorder and non-Hermiticity. For given non-Hermiticity corresponding to $\theta_L=i/2$, at weak disorder, states are (left) skin-localized, as effectively captured by $\langle \overline{\text{IPR}} \rangle$ $\simeq1/3$, and $\langle \overline{\text{mcom}} \rangle$ $\simeq0$. Increasing disorder initially detaches the boundary-concentrated skin modes, entering an intermediate NHSE-AL coexistence regime where eigenstates are neither sharply skin-localized (NHSE) nor yet sharply pinned to individual bulk sites (AL) --- reflected in a transient reduction of $\langle \overline{\text{IPR}} \rangle$. Further increase in disorder ultimately pins the eigenstates to isolated bulk sites, driving Anderson localization, supported by the recovery of $\langle \overline{\text{IPR}} \rangle$ toward unity and $\langle \overline{\text{mcom}} \rangle$ converging to $N/2$ ($\simeq 50$). Complementing these pictures, panels (b) and (e) demonstrate that increasing non-Hermiticity progressively drives the states toward boundary. The increasing $\langle \overline{\text{IPR}} \rangle$ captures the increasing localization strength due to increasing non-Hermiticity while decreasing $\langle \overline{\text{mcom}} \rangle$ reveals the accompanying directional skin accumulation. Finally, panels (c) and (f) illustrate the alternative route to realize such AL-NHSE crossover using magnetic field: for given non-Hermiticity and disorder strength, increasing magnetic field drives Anderson delocalization to give rise to an emergent NHSE, traversing an analogous transient dip in $\langle \overline{\text{IPR}} \rangle$ through the intermediate coexistence regime, as discussed above. This illustrates the phenomenon of \textit{magnetic field driven Anderson delocalization}.

The magnetic field thus acts as an additional tuning parameter that favors the AL$\to$NHSE crossover. This behavior stems from an effective suppression of the disorder strength due to the magnetic field induced inter-chain coupling (see Sec.~\ref{sec: mechanism underlying magnetic field driven Anderson delocalization} and Appendix~\ref{sec:appendix A,B,C} for a detailed discussion). Indeed, as shown in Fig.~\ref{fig:IPR,mcom}e, a finite magnetic field induces a clear directional bias in the eigenstates — reflected in $\langle \overline{\text{mcom}} \rangle \simeq 26$ at weak non-Hermiticity, despite the prevailing strong disorder ($W/J=8$) — whereas independently verifying the $B = 0$ case under the same parameters yields $\langle \overline{\text{mcom}} \rangle \simeq 50$, characteristic of Anderson-localized states symmetrically distributed across the bulk.

Fig.~\ref{fig:mcom heatmap} summarizes the triple interplay more directly. As depicted by Fig.~\ref{fig:mcom heatmap}(a,b), for a given non-Hermiticity, stronger disorder requires stronger magnetic field to induce Anderson delocalization. Put another way, for a given disorder strength, weaker non-Hermiticity necessitates a stronger magnetic field [Fig.~\ref{fig:mcom heatmap}(c,d)]. Most importantly, under a finite magnetic field strength, the familiar competition between disorder and non-Hermiticity --- as observed in the spinless framework [Fig.~\ref{fig:Representative NHSE, AL}] --- still persists here, but with modified boundaries [Fig.~\ref{fig:mcom heatmap}(e,f)]. An increase in the magnetic field translates the NHSE (red) regime rightwards, once again indicating that the AL$\rightarrow$NHSE crossover is favored.

\subsection{Additional remarks on the Magnetic field}
Before concluding the discussion of the numerical results, we make two additional observations regarding the role of the magnetic field beyond its central role in driving Anderson delocalization.

\begin{widetext}

\begin{figure}[t]
    \centering
    \includegraphics[width=0.9\linewidth]{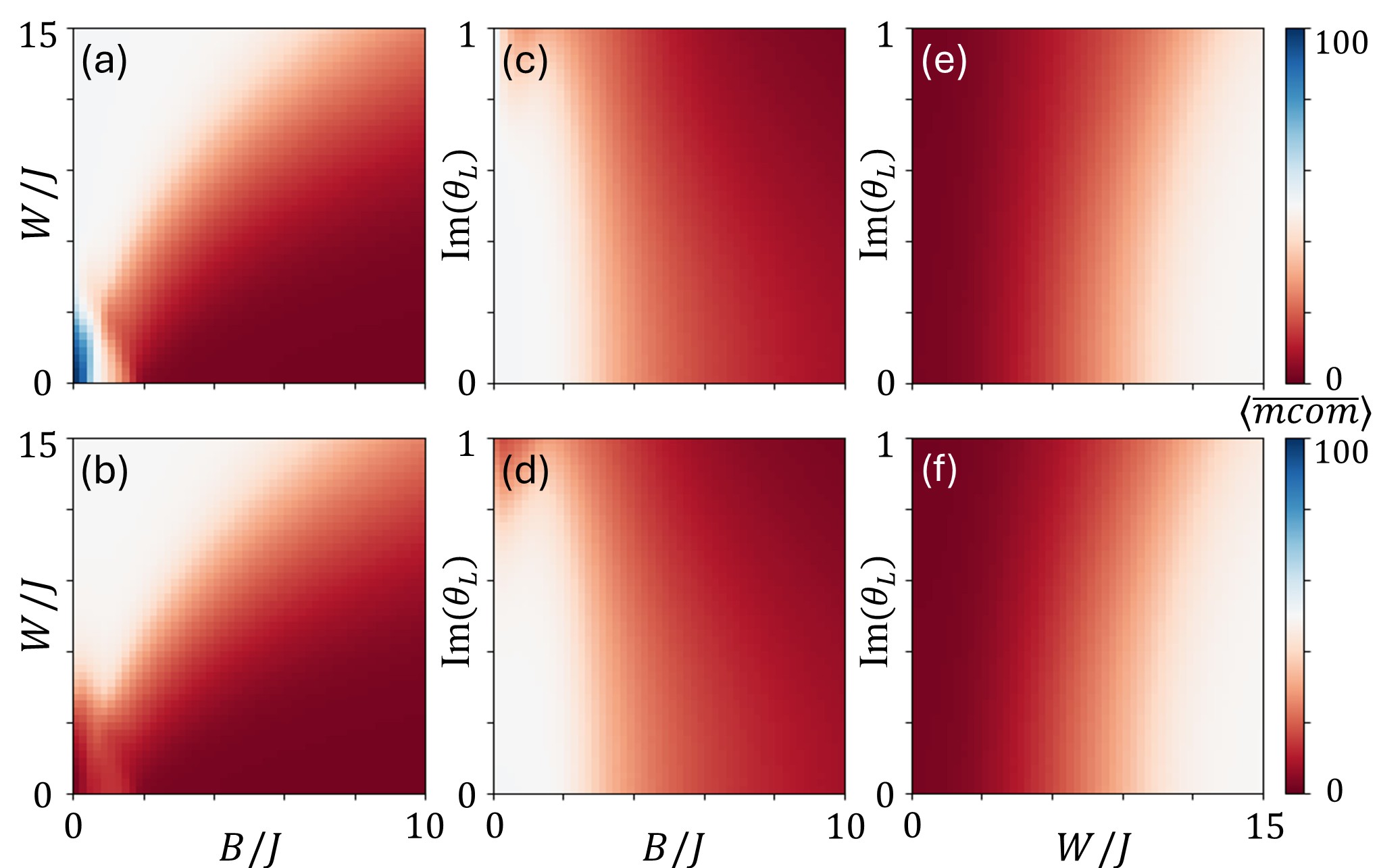}
    \caption{$\langle \overline{\text{mcom}} \rangle$ plots for both spin sectors having anti-symmetrically correlated disorders, represented in the \textbf{(a,b)} $B-W$ plane with $\theta_L=i/2$, \textbf{(c,d)} $\theta_L-B$ plane with $W/J=8$, and \textbf{(e,f)} $\theta_L-W$ plane with $B/J=4$. The upper (lower) panel represent up (down)-spin sector. The red (blue) region indicates left (right)-NHSE, and the white denotes Anderson localization (or extended states in limiting cases). The results are averaged over an ensemble of 1000 disorder realizations with $\theta_R=1$ and $N=100$.}
    \label{fig:mcom heatmap}
\end{figure}

\end{widetext}

\textit{Spin-polarized bidirectional NHSE:}
As seen in Figs.~\ref{fig:mcom heatmap}(a,b), at weak disorder and weak magnetic field, the two spin sectors exhibit opposite boundary localization: the up‑spin accumulates at the left edge (red), while the down‑spin accumulates at the right (blue). This spin‑polarized bidirectional skin effect originates from the opposite non‑reciprocities engineered by our gauge choice ($\theta_L\in\mathbb{C}$, $\theta_R\in\mathbb{R}$) in the decoupled chains. Increasing the magnetic field eventually aligns both sectors to the same direction, as also reported in our earlier work~\cite{sanahal1}. 

\textit{Magnetic-field-induced NHSE in reciprocal systems:}
A more subtle feature appears in Figs.~\ref{fig:mcom heatmap}(e,f) along the Im$(\theta_L)=0$ horizon. Here, even though the gauge fields are both real ($\theta_L,\theta_R\in\mathbb{R}$), the system exhibits NHSE (red region). Real gauge do not directly modulate hoping amplitudes unlike imaginary ones, and therefore cannot drive NHSE by themselves~\cite{areview,sanahal1}. However, under an external Zeeman field, NHSE emerges~\cite{spinful}. This magnetic field induced NHSE phenomenon originates from the magnetic field modulation of effective hoping amplitudes, generating effective non-reciprocity in the system. A detailed derivation is given in Appendix~\ref{sec:AppendixD}. Apart from this, the familiar competition between non-Hermiticity and disorder remains operative, with increasing disorder strength progressively driving the system toward the Anderson-localized (AL) regime (white region).

\section{The Underlying Physical Mechanism \label{sec: mechanism underlying magnetic field driven Anderson delocalization}}
The results in Sec.~\ref{sec: Results} demonstrate that an external in-plane magnetic field can drive Anderson delocalization and induce non-Hermitian skin modes in strongly disordered spinful non-Hermitian systems. In this section, we elucidate the physical mechanism underlying this transition. We show that the delocalization transition can be understood through a effective suppression of disorder strength mediated by inter-chain coupling that arises in our two-chain system due to the in-plane magnetic field.

\subsection{Inter-chain coupling driven Anderson delocalization \label{sub sec: Restoration of suppressed non-Hermiticity via inter-chain coupling}}
To begin with, we transform the Hamiltonian featuring anti-symmetric disorder configuration into the basis:

\begin{equation*}
        \begin{bmatrix}
        \alpha_{n,a} \\
        \alpha_{n,b}
    \end{bmatrix}
    = 
    \begin{bmatrix}
        \text{cos}(\phi_n) & \text{sin}(\phi_n) \\
        -\text{sin}(\phi_n) & \text{cos}(\phi_n) 
    \end{bmatrix}
        \begin{bmatrix}
        c_{n,\uparrow} \\
        c_{n,\downarrow}
    \end{bmatrix}
\end{equation*}
In this representation, the transformed Hamiltonian maps onto an effective Creutz ladder~\cite{Michael_Creutz}, consisting of two chains --- $a$ and $b$ --- that feature non‑reciprocal hoppings dictated by the gauge fields, and effective disordered onsite potentials of the form $\pm (\Delta_n^2+B^2)^{1/2}$ on the two chains [see Appendix~\ref{sec:AppendixA}]. The effective onsite potential receives contributions from both the $c_{n,s}^\dagger c_{n,s}$ and $c_{n,s}^\dagger c_{n,s'}$ ($s\neq s'$) terms. While the former corresponds to the disordered potentials $\Delta_n^{{\uparrow}}$ and $\Delta_n^{{\downarrow}}$ in the original spin chains, the latter represents onsite inter-chain coupling between them, which, in our context, is provided by the in-plane magnetic field $B$.

\begin{figure}
    \centering
    \includegraphics[width=1\linewidth]{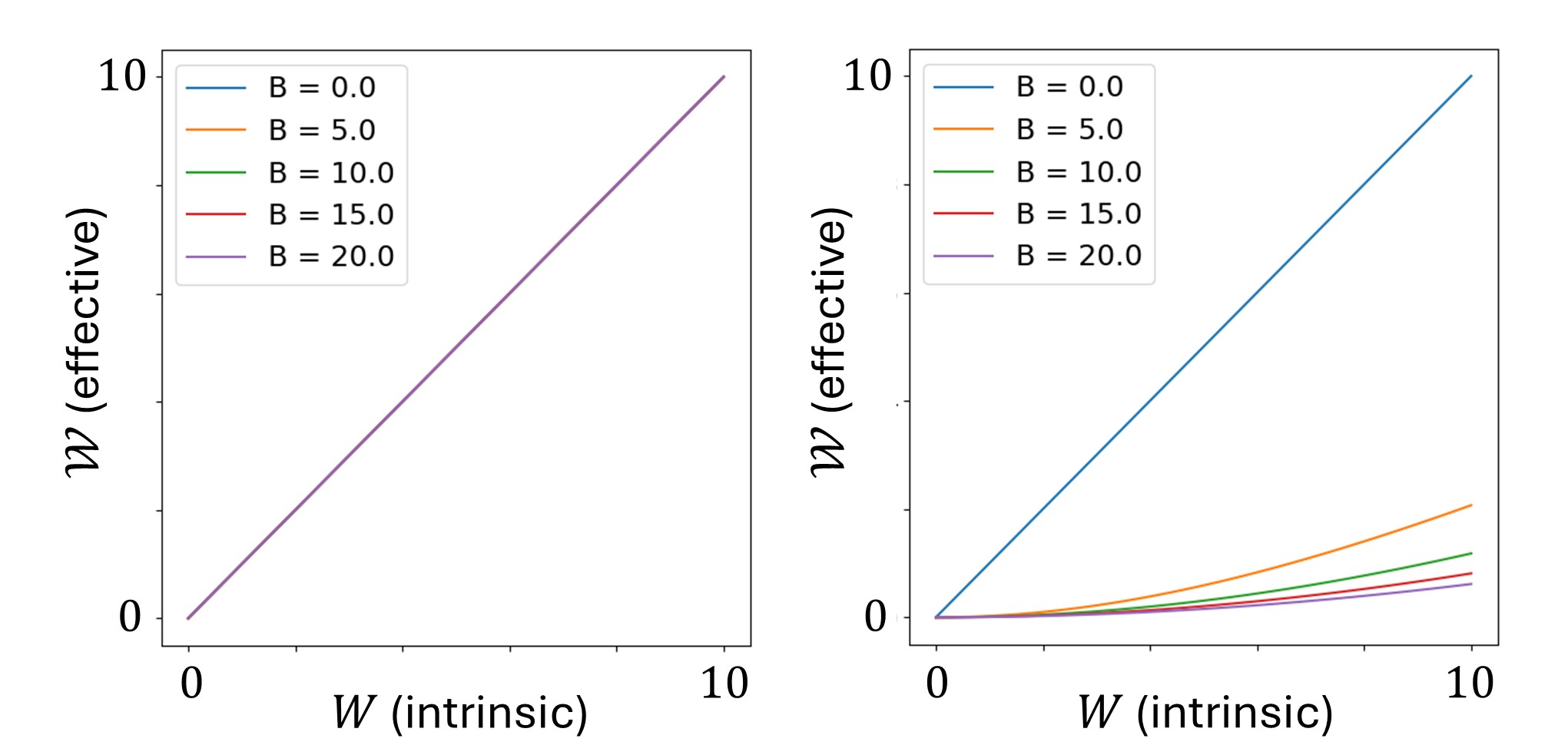}
    \caption{ Effective versus intrinsic disorder strength under varying magnetic fields when disorders in the two chains are \textbf{(a)} symmetrically, and \textbf{(b)} anti-symmetrically correlated. While the magnetic field does not affect the disorder strength in the symmetric case, it suppresses the disorder to yield a weaker effective disorder strength in the system under anti-symmetrically correlated disorder. }
    \label{fig: effective verrsus intrinsic disorder strength}
\end{figure}

As $\Delta_n$ is uniformly distributed in $\left[ -W/2,W/2 \right]$, the effective onsite potentials in the two chains ($a,b$) of the ladder range in  
$$ 
\left[ B , \sqrt{B^2 + \frac{W^2}{4}} \right], 
\quad 
\left[ -\sqrt{B^2 + \frac{W^2}{4}} , -B \right] 
$$ 
respectively [see Appendix~\ref{sec:AppendixB}]. For any finite $B>0$, the effective disorder strength $\mathcal{W} = \sqrt{B^2 + W^2/4} \ - B$ is lesser than the intrinsic disorder strength $W$. Moreover, it decreases even further as $B$ increases [Fig.~\ref{fig: effective verrsus intrinsic disorder strength}]. This progressive suppression of the effective disorder allows the intrinsic non‑reciprocity to dominate, thereby driving the system from an Anderson‑localized regime into the non‑Hermitian skin regime. Thus, the magnetic field induced suppression of the effective disorder strength explains the Anderson delocalization observed in Sec.~\ref{sec: Results}.

It is instructive to note that the observed magnetic field driven Anderson delocalization is purely a non‑Hermitian phenomenon. In the Hermitian limit, although the magnetic field still suppresses the effective disorder, the absence of intrinsic non-reciprocity keeps the system strictly Anderson localized for any infinitesimal non‑zero disorder strength (see Appendix~\ref{sec:AppendixC}).

\section{Discussions and Conclusion \label{sec: Conclusion}} 
In summary, building on the established competition between disorder and non-Hermiticity, we uncovered how an external magnetic field mediates this interplay in a spinful extension of one-dimensional non-Hermitian systems. We have established that a magnetic field can serve as a controllable switch for driving Anderson delocalization in such systems, and mapped the resulting triple interplay among \textit{disorder}, \textit{non-Hermiticity}, and \textit{magnetic field} using \textit{inverse participation ratio} and the \textit{mean center of mass}. By carefully engineering antisymmetric disorder correlations across spin sectors, the magnetic field progressively suppresses the effective disorder width via a nonlinear transformation. This reduction allows the intrinsic non‑reciprocity, encoded in the complex gauge fields to dominate, driving Anderson‑localized states into directional skin modes. Crucially, this magnetic field driven delocalization is a purely non‑Hermitian phenomenon: in the Hermitian limit, despite identical disorder suppression, the system remains strictly Anderson localized, with no preferred boundary accumulation due to the absence of non‑reciprocity.

Our work establishes the magnetic field as a third control parameter --- alongside disorder and non-Hermiticity --- for engineering localization transitions, thereby opening new avenues for exploring rich triple interplay in experimentally accessible synthetic platforms where spin degrees and synthetic gauge fields are readily available. 

\section*{Acknowledgments}
We acknowledge the computing resources of `PARAM SHAVAK' at Computational Condensed Matter Physics Lab, Department of Physics, NIT Silchar. S.~N. acknowledges financial support from Anusandhan National Research Foundation (ANRF), Government of India via the Prime Minister's Early Career Research Grant: ANRF/ECRG/2024/005947/PMS.

\appendix
\renewcommand{\thesection}{\Alph{section}}
\section{MAGNETIC FIELD CONTROLLED ANDERSON DELOCALIZATION \label{sec:appendix A,B,C}}
\renewcommand{\theequation}{\thesection.\arabic{equation}}
\setcounter{equation}{0}

\subsection{Basis transformation and the Transformed Hamiltonian \label{sec:AppendixA}}
The generic Hamiltonian in Eqn.~\ref{eqn: spinful HN model}, expressed explicitly under our choice of gauge $\sigma_L,\sigma_R=\sigma_z$ and $\textbf{B}=B\hat{i}$ reads 
\begin{align}
    H = & \sum_{n=1}^{N-1} (Je^{i\theta_L\sigma_{z}} c_{n}^\dagger c_{n+1} + J e^{i\theta_R\sigma_{z}} c_{n+1}^\dagger c_{n})\ + \nonumber \\
    & \sum_{n=1}^{N} (\Delta_n^{(\uparrow)} c_{n,\uparrow}^{\dagger}c_{n,\uparrow} + \Delta_n^{(\downarrow)} c_{n,\downarrow}^{\dagger}c_{n,\downarrow}) + \sum_{n=1}^{N} (B\sigma_x) c_n^{\dagger}c_n . 
    \label{eqn: spinful Hamiltonian for zzx}
\end{align}

\subsubsection{Case: Anti-symmetrically correlated disorder}
Starting with the case of anti-symmetrically correlated disorder where $\Delta_n^{(\uparrow)} = -\Delta_n^{(\downarrow)}$, we re-express the above Hamiltonian as
\begin{align}
    H_- = & \sum_{n=1}^{N-1} (Je^{i\theta_L\sigma_{z}} c_{n}^\dagger c_{n+1} + J e^{i\theta_R\sigma_{z}} c_{n+1}^\dagger c_{n})\ + \nonumber \\
    & \sum_{n=1}^{N} (\Delta_n \sigma_z + B\sigma_x) c_n^{\dagger}c_n.
    \label{eqn: original anti-symmetric}
\end{align}
where, the first two terms in the Hamiltonian represent kinetic contributions, whereas the last term indicate onsite potentials, yielding the form $H_-=H^{hop} + H_-^{onsite}$.

Our prior objective is to diagonalize the $H_-^{onsite}$ for reasons that will soon be clear. For that, we perform a unitary transformation via change of basis 
\begin{align*}
    \begin{bmatrix}
        c_{n,\uparrow} \\
        c_{n,\downarrow}
    \end{bmatrix}
    \longrightarrow
    \begin{bmatrix}
        \text{cos}(\phi_n) & -\text{sin}(\phi_n) \\
        \text{sin}(\phi_n) & \text{cos}(\phi_n) 
    \end{bmatrix}
    \begin{bmatrix}
        \alpha_{n,a} \\
        \alpha_{n,b}
    \end{bmatrix}
\end{align*}
such that Eqn.~\ref{eqn: original anti-symmetric} transforms as

\begin{widetext}

    \begin{align}
        \mathscr{H_-} = & \sum_{n=1}^{N-1} \ (U_n^\dagger Je^{i\theta_L\sigma_{z}} U_{n+1}) \alpha_{n}^\dagger \alpha_{n+1} + \sum_{n=1}^{N-1} \ (U_{n+1}^\dagger Je^{i\theta_R\sigma_{z}} U_{n}) \alpha_{n+1}^\dagger \alpha_{n} \ + \sum_{n=1}^{N} (U_n^\dagger(\Delta_n \sigma_z + B\sigma_x)U_n) \alpha_n^{\dagger} \alpha_n.
        \label{eqn: transformed anti-symmetric}
    \end{align}
    where $U_n$ is the site-dependent $2\times2$ transformation matrix. The site-dependence of $U_n$ arises from the fact that the disordered onsite potentials $\Delta_n$ in the Hamiltonian (Eqn.~\ref{eqn: original anti-symmetric}) vary with sites. The $2\times2$ matrices in each term in Eqn.~\ref{eqn: transformed anti-symmetric} i.e. $U_n^\dagger Je^{i\theta_L\sigma_{z}} U_{n+1}$, $U_{n+1}^\dagger Je^{i\theta_R\sigma_{z}} U_{n}$, and $U_n^\dagger(\Delta_n \sigma_z + B\sigma_x)U_n$ when expressed explicitly read \\
    
    \begin{equation}
    \begin{bmatrix}
        Je^{i\theta_L} \text{cos}(\phi_n) \text{cos}(\phi_{n+1}) + Je^{-i\theta_L} \text{sin}(\phi_n) \text{sin}(\phi_{n+1}) & 
        -Je^{i\theta_L} \text{cos}(\phi_n) \text{sin}(\phi_{n+1}) + Je^{-i\theta_L} \text{sin}(\phi_n) \text{cos}(\phi_{n+1}) \\
        -Je^{i\theta_L} \text{sin}(\phi_n) \text{cos}(\phi_{n+1}) + Je^{-i\theta_L} \text{cos}(\phi_n) \text{sin}(\phi_{n+1}) & 
        Je^{i\theta_L} \text{sin}(\phi_n) \text{sin}(\phi_{n+1}) + Je^{-i\theta_L} \text{cos}(\phi_n) \text{cos}(\phi_{n+1})
    \end{bmatrix}, 
    \label{exp: transformed left hoping}
    \end{equation}

    \begin{equation}
    \begin{bmatrix}
        Je^{i\theta_R} \text{cos}(\phi_{n+1}) \text{cos}(\phi_{n}) + Je^{-i\theta_R} \text{sin}(\phi_{n+1}) \text{sin}(\phi_{n}) & 
        -Je^{i\theta_R} \text{cos}(\phi_{n+1}) \text{sin}(\phi_{n}) + Je^{-i\theta_R} \text{sin}(\phi_{n+1}) \text{cos}(\phi_{n}) \\
        -Je^{i\theta_R} \text{sin}(\phi_
        {n+1}) \text{cos}(\phi_{n}) + Je^{-i\theta_R} \text{cos}(\phi_{n+1}) \text{sin}(\phi_{n}) & 
        Je^{i\theta_R} \text{sin}(\phi_{n+1}) \text{sin}(\phi_{n}) + Je^{-i\theta_R} \text{cos}(\phi_{n+1}) \text{cos}(\phi_{n})
    \end{bmatrix},
    \label{exp: transformed right hoping}
    \end{equation}
    
    and
    
    \begin{equation}
    \begin{bmatrix}
        \sqrt{\Delta_n^2+B^2} & 0 \\
        0 & - \sqrt{\Delta_n^2+B^2}
    \end{bmatrix},
    \label{exp: transformed onsite term}
    \end{equation}
    respectively, with 
    \begin{equation}
        \text{tan}(2\phi_n) = \frac{B}{\Delta_n}.   
        \label{exp: tan(phi)}
    \end{equation}
    
\end{widetext}

In the new basis $[\alpha_{n,a}, \alpha_{n,b}]^T$, the transformed Hamiltonian takes the form 
$$
\mathscr{H}_- = \mathscr{H}^{hop} + \mathscr{H}_-^{onsite}
$$
where $\mathscr{H}^{hop}$ consists of the first two terms in Eqn.~\ref{eqn: transformed anti-symmetric}, arising from $H_-^{hop}$, and $\mathscr{H}_-^{onsite}$ represents the last term, arising from $H_-^{onsite}$. While $\mathscr{H}_-^{onsite}$ depends on $\Delta_n$ and $B$ (see expression~\ref{exp: transformed onsite term}) which is quite expected, the $\mathscr{H}^{hop}$ interestingly depends on all parameters: $\Delta_n$ and $B$ via $\phi_n$, and $\theta_{L,R}$ (see expression~\ref{exp: transformed left hoping} and \ref{exp: transformed right hoping}).

Physically, the $\mathscr{H}^{hop}$ resembles a Creutz ladder consisting of two chains --- $a$ and $b$ --- with hopping strengths dependent on all $\theta_{L,R},\Delta_n$, and $B$. For the case: $\theta_L\in\mathbb{C}$, and $\theta_R\in\mathbb{R}$, we clearly see, the hoping amplitudes are modulated unequally. In particular, $|Je^{i\theta_L}|\rightarrow |Je^{-\text{Im}(\theta_L)}|$ and $|Je^{-i\theta_L}|\rightarrow |Je^{\text{Im}(\theta_L)}|$ while $|Je^{\pm i\theta_R}|$ remains $1$. This gauge field-induced non-reciprocity results to the competition between AL and NHSE. 

On the other hand, the $\mathscr{H}_-^{onsite}$ represents disordered effective onsite potentials on the $a$, $b$ chains of the ladder 
\begin{equation*}
\mathscr{H}_-^{onsite} = 
    \begin{bmatrix}
        \sqrt{\Delta_n^2+B^2} & 0 \\
        0 & -\sqrt{\Delta_n^2+B^2}
    \end{bmatrix},
\end{equation*}
where $\Delta_n$ varies uniformly in $[-W/2,W/2]$.

\subsubsection{Case: Symmetrically correlated disorder}
In the case of symmetrically correlated disorder where $\Delta_n^{(\uparrow)} = \Delta_n^{(\downarrow)}$, we re-express Eqn.~\ref{eqn: spinful Hamiltonian for zzx} as
\begin{align}
    H_+ = & \sum_{n=1}^{N-1} (Je^{i\theta_L\sigma_{z}} c_{n}^\dagger c_{n+1} + J e^{i\theta_R\sigma_{z}} c_{n+1}^\dagger c_{n})\ + \nonumber \\
    & \sum_{n=1}^{N} (\Delta_n \mathbb{I} + B\sigma_x) c_n^{\dagger}c_n.
\end{align}

Following similar approach as in the case of anti-symmetric disorder, here we get 
$$
\mathscr{H}_+ = \mathscr{H}^{hop} + \mathscr{H}_+^{onsite}
$$
where 
\begin{equation*}
\mathscr{H}_+^{onsite} = 
    \begin{bmatrix}
        \Delta_n+B & 0 \\
        0 & \Delta_n-B
    \end{bmatrix},
\end{equation*}
where $\Delta_n$ varies uniformly in $[-W/2,W/2]$.

\subsection{Magnetic field suppressed effective disorder strength \label{sec:AppendixB}}
Under anti-symmetrically correlated disorder, the ladder hosts effective disordered onsite potentials of the form: $\pm (\Delta_n^2+B^2)^{1/2}$ in the $a,b$ chain respectively. As the random potentials $\Delta_n$ are uniformly distributed in $[-W/2,\,W/2]$, the values of the effective potentials, in $a$-chain for instance, consequently range from $B$ (when $\Delta_n = 0$) to $\sqrt{B^2 + W^2/4}$ (when $\Delta_n = \pm W/2$). The effective disorder strength therefore becomes
\begin{equation}
    \mathcal{W}(B,W) = \sqrt{B^2 + W^2/4} \ - B. 
    \label{eqn:suppression}
\end{equation}
For any finite magnetic field $B>0$, this width $\mathcal{W}$ is strictly smaller than the intrinsic disorder strength $W$. Moreover, it decreases monotonically as $B$ increases, as illustrated in Fig.~\ref{fig: effective verrsus intrinsic disorder strength} in the main text. This progressive suppression of the effective disorder strength under anti-symmetric disorder results to increasing manifestation of the intrinsic non-reciprocity in the system. Hence the magnetic field promotes the AL$\rightarrow$NHSE crossover. 

In strong field limit where $B>W$, the description becomes even more explicit. As Eqn.~\ref{eqn:suppression} reduces to 
$$ 
\mathcal{W}(B,W) = \frac{W^2}{8B},
$$ 
we see increasing $B$ progressively decreases the $\mathcal{W}$.

Meanwhile, under symmetric disorder, the effective onsite potentials are of the form $(\Delta_n \pm B)$. Unlike above, since $\Delta_n\in [-W/2,W/2]$, the fluctuation width here remains the same $(W)$ in both the chains; the magnetic field merely shifts the entire distribution by $B$ without suppressing its width. For instance, in $\alpha$ chain, the onsite potentials range in 
\begin{equation*}
    \left[ \left( -\frac{W}{2}+B \right), \left( \frac{W}{2}+B\right) \right]. 
\end{equation*}

Thus, under symmetrically correlated disorder, the magnetic field does not suppress the disorder strength, and therefore is unable to drive the Anderson delocalization phenomenon.

\subsection{Absence of Anderson delocalization in the Hermitian limit \label{sec:AppendixC}}
In the Hermitian limit $\theta_R=-\theta_L$ ($\theta_{L,R}\in \mathbb{R}$), $\mathscr{H}_-^{\dagger}=\mathscr{H}_-$. Because $U_n^\dagger J e^{i\theta_L\sigma_z}U_{n+1}$ and $U_{n+1}^\dagger J e^{i\theta_R\sigma_z}U_n$ become Hermitian conjugates of each other, the hopping amplitudes are reciprocal (this can also be verified directly from the explicit matrix expressions in Eqs.~\ref{exp: transformed left hoping} and \ref{exp: transformed right hoping}).

Even though the magnetic field suppresses the effective disorder strength $\mathcal{W}$ as derived before, the complete absence of non‑reciprocity in the hopping prevents the directional bias needed for the NHSE. Consequently, for any infinitesimal disorder $\mathcal{W}>0$, the system remains strictly Anderson localized. Thus the magnetic‑field‑driven delocalization observed in the main text is inherently a non‑Hermitian phenomenon, relying on the competition between suppressed disorder and intrinsic non‑reciprocal hopping. 

\section{MAGNETIC FIELD INDUCED NHSE IN A RECIPROCAL NON-HERMITIAN CHAIN \label{sec:AppendixD}}
When the gauge fields are real, i.e., $\theta_L,\theta_R\in\mathbb{R}$, the bare hoppings satisfy $|Je^{i\theta_L}|=|Je^{i\theta_R}|$ and $|Je^{-i\theta_L}|=|Je^{-i\theta_R}|$. Hence the system is reciprocal and does not exhibit the NHSE in itself~\cite{areview}. However, as shown in prior works~\cite{sanahal1,spinful}, applying an external in-plane magnetic field $B$ can induce NHSE even in such reciprocal systems. The origin of this phenomenon uncovers after the site-dependent unitary transformation presented in Appendix~\ref{sec:AppendixA}.

For the antisymmetric disorder case, the transformed Hamiltonian Eqn.~\ref{eqn: transformed anti-symmetric} ($\mathscr{H}_-^{hop}$ to be specific) contains hopping matrices $U_n^\dagger J e^{i\theta_L\sigma_z} U_{n+1}$ (leftward) and $U_{n+1}^\dagger J e^{i\theta_R\sigma_z} U_n$ (rightward) that depend on all $\theta_{L,R}, \Delta_n$ and $B$. Even though $\theta_L,\theta_R$ are real, the rotation angles $\phi_n$ (determined by $\tan(2\phi_n) = B/\Delta_n$) introduce an effective magnetic field dependence in the hopping amplitudes. Explicitly, from expressions~\ref{exp: transformed left hoping} and \ref{exp: transformed right hoping}, the magnitudes of the matrix elements representing leftward, rightward hoppings are no longer necessarily symmetric (though not exactly traceable), This effectively breaks the reciprocity and the system consequently develops a directional bias, generating boundary accumulation of eigenstates. 

To better elucidate, we demonstrate the limiting case of $B=0$. Using Eqn.~\ref{exp: tan(phi)} under this limit ($\phi=0$), Eqn.~\ref{eqn: transformed anti-symmetric} reduces to

\begin{widetext}

\begin{align*}
    \mathscr{H}_- = \sum_{n=1}^{N-1}
    \begin{bmatrix}
        Je^{i\theta_L} & 0 \\
        0 & Je^{-i\theta_L}
    \end{bmatrix}
    \alpha_{n}^\dagger \alpha_{n+1}
    +
    \sum_{n=1}^{N-1}
    \begin{bmatrix}
        Je^{i\theta_R} & 0 \\
        0 & Je^{-i\theta_R}
    \end{bmatrix}
    \alpha_{n+1}^\dagger \alpha_n
    + 
    \sum_{n=1}^{N}
    \begin{bmatrix}
        \Delta_n & 0 \\
        0 & -\Delta_n
    \end{bmatrix}
    \alpha_{n}^\dagger \alpha_n.
\end{align*}

\end{widetext}

which is identical to the original Hamiltonian. Clearly, the hopping strengths are only gauge-dependent, exhibiting reciprocity: $|Je^{i\theta_L}|=|Je^{i\theta_R}|$ and $|Je^{-i\theta_L}|=|Je^{-i\theta_R}|$. Notice as we turn $B$ on, the hoppings become $B$-dependent, and are no longer necessarily reciprocal. This effect is purely magnetic field induced and coexists with the disorder suppression mechanism.


\bibliographystyle{unsrt}
\bibliography{Ref}

@article{topologicalorigin,
  title={Topological origin of non-Hermitian skin effects},
  author={Okuma, Nobuyuki and Kawabata, Kohei and Shiozaki, Ken and Sato, Masatoshi},
  journal={Physical review letters},
  volume={124},
  number={8},
  pages={086801},
  year={2020},
  publisher={APS}
}

@article{anomalousedgestates,
  title={Anomalous edge state in a non-Hermitian lattice},
  author={Lee, Tony E},
  journal={Physical review letters},
  volume={116},
  number={13},
  pages={133903},
  year={2016},
  publisher={APS}
}

@article{principles&prospects,
  title={Non-Hermitian topological phases: principles and prospects},
  author={Banerjee, Ayan and Sarkar, Ronika and Dey, Soumi and Narayan, Awadhesh},
  journal={Journal of Physics: Condensed Matter},
  volume={35},
  number={33},
  pages={333001},
  year={2023},
  publisher={IOP Publishing},
}

@article{bender1998,
  title={Real spectra in non-Hermitian Hamiltonians having P T symmetry},
  author={Bender, Carl M and Boettcher, Stefan},
  journal={Physical review letters},
  volume={80},
  number={24},
  pages={5243},
  year={1998},
  publisher={APS}
}

@article{topPhasesofNHsystem,
  title={Topological phases of non-Hermitian systems},
  author={Gong, Zongping and Ashida, Yuto and Kawabata, Kohei and Takasan, Kazuaki and Higashikawa, Sho and Ueda, Masahito},
  journal={Physical Review X},
  volume={8},
  number={3},
  pages={031079},
  year={2018},
  publisher={APS}
}

@article{areview,
  title={A review on non-Hermitian skin effect},
  author={Zhang, Xiujuan and Zhang, Tian and Lu, Ming-Hui and Chen, Yan-Feng},
  journal={Advances in Physics: X},
  volume={7},
  number={1},
  pages={2109431},
  year={2022},
  publisher={Taylor \& Francis}
}

@article{spinful,
  title={Skin effect in non-Hermitian systems with SU (2) gauge fields},
  author={Zhang, Wenna and Hu, Yutao and Zhang, Hongyi and Liu, Xiang and Shen, Yuecheng and Veronis, Georgios and Al{\`u}, Andrea and Huang, Yin and Luo, Wenchen},
  journal={Physical Review B},
  volume={112},
  number={12},
  pages={125164},
  year={2025},
  publisher={APS}
}

@article{correspondence,
  title={Correspondence between winding numbers and skin modes in non-Hermitian systems},
  author={Zhang, Kai and Yang, Zhesen and Fang, Chen},
  journal={Physical Review Letters},
  volume={125},
  number={12},
  pages={126402},
  year={2020},
  publisher={APS}
}

@article{aperspective,
  title={The non-Hermitian skin effect: A perspective},
  author={Gohsrich, Julius T and Banerjee, Ayan and Kunst, Flore K},
  journal={arXiv preprint arXiv:2410.23845},
  year={2024}
}

@article{NH_topological_phenomena:A_review,
  title={Non-Hermitian topological phenomena: A review},
  author={Okuma, Nobuyuki and Sato, Masatoshi},
  journal={Annual Review of Condensed Matter Physics},
  volume={14},
  number={1},
  pages={83--107},
  year={2023},
  publisher={Annual Reviews}
}

@article{PT_sym_photonic_lattice,
  title={Parity--time synthetic photonic lattices},
  author={Regensburger, Alois and Bersch, Christoph and Miri, Mohammad-Ali and Onishchukov, Georgy and Christodoulides, Demetrios N and Peschel, Ulf},
  journal={Nature},
  volume={488},
  number={7410},
  pages={167--171},
  year={2012},
  publisher={Nature Publishing Group UK London}
}

@article{PT_sym_microring_lasers,
  title={Parity-time--symmetric microring lasers},
  author={Hodaei, Hossein and Miri, Mohammad-Ali and Heinrich, Matthias and Christodoulides, Demetrios N and Khajavikhan, Mercedeh},
  journal={Science},
  volume={346},
  number={6212},
  pages={975--978},
  year={2014},
  publisher={American Association for the Advancement of Science}
}

@article{PT_sym_whispering_gallery_microcavities,
  title={Parity--time-symmetric whispering-gallery microcavities},
  author={Peng, Bo and {\"O}zdemir, {\c{S}}ahin Kaya and Lei, Fuchuan and Monifi, Faraz and Gianfreda, Mariagiovanna and Long, Gui Lu and Fan, Shanhui and Nori, Franco and Bender, Carl M and Yang, Lan},
  journal={Nature Physics},
  volume={10},
  number={5},
  pages={394--398},
  year={2014},
  publisher={Nature Publishing Group UK London}
}

@article{nonAbelian_effects_in_dissipative_photonics,
  title={Non-Abelian effects in dissipative photonic topological lattices},
  author={Parto, Midya and Leefmans, Christian and Williams, James and Nori, Franco and Marandi, Alireza},
  journal={Nature Communications},
  volume={14},
  number={1},
  pages={1440},
  year={2023},
  publisher={Nature Publishing Group UK London}
}

@article{Observing_PTsym_in_optics,
  title={Observation of parity--time symmetry in optics},
  author={R{\"u}ter, Christian E and Makris, Konstantinos G and El-Ganainy, Ramy and Christodoulides, Demetrios N and Segev, Mordechai and Kip, Detlef},
  journal={Nature physics},
  volume={6},
  number={3},
  pages={192--195},
  year={2010},
  publisher={Nature Publishing Group UK London}
}

@article{EPs_enhance_sensing_in_optical_microcavity,
  title={Exceptional points enhance sensing in an optical microcavity},
  author={Chen, Weijian and Kaya {\"O}zdemir, {\c{S}}ahin and Zhao, Guangming and Wiersig, Jan and Yang, Lan},
  journal={Nature},
  volume={548},
  number={7666},
  pages={192--196},
  year={2017},
  publisher={Nature Publishing Group UK London}
}

@article{NH_photonics_based_on_PTsym,
  title={Non-Hermitian photonics based on parity--time symmetry},
  author={Feng, Liang and El-Ganainy, Ramy and Ge, Li},
  journal={Nature Photonics},
  volume={11},
  number={12},
  pages={752--762},
  year={2017},
  publisher={Nature Publishing Group}
}

@article{GDSE_in_reciprocal_photonic,
  title={Geometry-dependent skin effects in reciprocal photonic crystals},
  author={Fang, Zhening and Hu, Mengying and Zhou, Lei and Ding, Kun},
  journal={Nanophotonics},
  volume={11},
  number={15},
  pages={3447--3456},
  year={2022},
  publisher={De Gruyter}
}

@article{Gen_BBC_in_NH_topoelectric_circuit,
  title={Generalized bulk--boundary correspondence in non-Hermitian topolectrical circuits},
  author={Helbig, Tobias and Hofmann, Tobias and Imhof, S and Abdelghany, M and Kiessling, T and Molenkamp, LW and Lee, CH and Szameit, A and Greiter, M and Thomale, R},
  journal={Nature Physics},
  volume={16},
  number={7},
  pages={747--750},
  year={2020},
  publisher={Nature Publishing Group UK London}
}

@article{nonAbelian_gauge_fields_in_circuit,
  title={Non-Abelian gauge fields in circuit systems},
  author={Wu, Jiexiong and Wang, Zhu and Biao, Yuanchuan and Fei, Fucong and Zhang, Shuai and Yin, Zepeng and Hu, Yejian and Song, Ziyin and Wu, Tianyu and Song, Fengqi and others},
  journal={Nature Electronics},
  volume={5},
  number={10},
  pages={635--642},
  year={2022},
  publisher={Nature Publishing Group UK London}
}

@article{HOSE_in_NH_topoelectric_circuit,
  title={Observation of hybrid higher-order skin-topological effect in non-Hermitian topolectrical circuits},
  author={Zou, Deyuan and Chen, Tian and He, Wenjing and Bao, Jiacheng and Lee, Ching Hua and Sun, Houjun and Zhang, Xiangdong},
  journal={Nature Communications},
  volume={12},
  number={1},
  pages={7201},
  year={2021},
  publisher={Nature Publishing Group UK London}
}

@article{observation_of_SDBE_in_NH_electric,
  title={Observation of size-dependent boundary effects in non-Hermitian electric circuits},
  author={Su, Luhong and Guo, Cui-Xian and Wang, Yongliang and Li, Li and Ruan, Xinhui and Du, Yanjing and Chen, Shu and Zheng, Dongning},
  journal={Chinese Physics B},
  volume={32},
  number={3},
  pages={038401},
  year={2023},
  publisher={IOP Publishing}
}

@article{ReciprocalSE_and_its_realization,
  title={Reciprocal skin effect and its realization in a topolectrical circuit},
  author={Hofmann, Tobias and Helbig, Tobias and Schindler, Frank and Salgo, Nora and Brzezi{\'n}ska, Marta and Greiter, Martin and Kiessling, Tobias and Wolf, David and Vollhardt, Achim and Kaba{\v{s}}i, Anton and others},
  journal={Physical review research},
  volume={2},
  number={2},
  pages={023265},
  year={2020},
  publisher={APS}
}

@article{NHSE_in_a_NH_electrical_circuit,
  title={Non-Hermitian skin effect in a non-Hermitian electrical circuit},
  author={Liu, Shuo and Shao, Ruiwen and Ma, Shaojie and Zhang, Lei and You, Oubo and Wu, Haotian and Xiang, Yuan Jiang and Cui, Tie Jun and Zhang, Shuang},
  journal={Research},
  year={2021},
  publisher={AAAS}
}

@article{DYnamic_signs_of_NHSE_in_UCatoms,
  title={Dynamic signatures of non-Hermitian skin effect and topology in ultracold atoms},
  author={Liang, Qian and Xie, Dizhou and Dong, Zhaoli and Li, Haowei and Li, Hang and Gadway, Bryce and Yi, Wei and Yan, Bo},
  journal={Physical review letters},
  volume={129},
  number={7},
  pages={070401},
  year={2022},
  publisher={APS}
}

@article{2D_NHSE_in_UCfermigas,
  title={Two-dimensional non-Hermitian skin effect in an ultracold Fermi gas},
  author={Zhao, Entong and Wang, Zhiyuan and He, Chengdong and Poon, Ting Fung Jeffrey and Pak, Ka Kwan and Liu, Yu-Jun and Ren, Peng and Liu, Xiong-Jun and Jo, Gyu-Boong},
  journal={Nature},
  volume={637},
  number={8046},
  pages={565--573},
  year={2025},
  publisher={Nature Publishing Group UK London}
}

@article{theoretical_prediction_of_NHSE_in_UCatom,
  title={Theoretical prediction of a non-Hermitian skin effect in ultracold-atom systems},
  author={Guo, Sibo and Dong, Chenxiao and Zhang, Fuchun and Hu, Jiangping and Yang, Zhesen},
  journal={Physical Review A},
  volume={106},
  number={6},
  pages={L061302},
  year={2022},
  publisher={APS}
}

@article{NHSE_in_periodically_driven_dissipative_UCatom,
  title={Non-Hermitian skin effect in periodically driven dissipative ultracold atoms},
  author={Cai, Zhao-Fan and Liu, Tao and Yang, Zhongmin},
  journal={Physical Review A},
  volume={109},
  number={6},
  pages={063329},
  year={2024},
  publisher={APS}
}

@article{Observation_of_NHSE-in_mechanical_metamaterial,
  title={Observation of non-Hermitian topology and its bulk--edge correspondence in an active mechanical metamaterial},
  author={Ghatak, Ananya and Brandenbourger, Martin and Van Wezel, Jasper and Coulais, Corentin},
  journal={Proceedings of the National Academy of Sciences},
  volume={117},
  number={47},
  pages={29561--29568},
  year={2020},
  publisher={National Academy of Sciences}
}

@article{Phy_of_open_QM_systems,
  title={A non-Hermitian Hamilton operator and the physics of open quantum systems},
  author={Rotter, Ingrid},
  journal={Journal of Physics A: Mathematical and Theoretical},
  volume={42},
  number={15},
  pages={153001},
  year={2009},
  publisher={IOP Publishing}
}

@article{NHSE_in_open_QM_systems,
  title={Non-Hermitian skin effect and chiral damping in open quantum systems},
  author={Song, Fei and Yao, Shunyu and Wang, Zhong},
  journal={arXiv preprint arXiv:1904.08432},
  year={2019}
}

@article{Liouvillian_SE_in_exactly_solvable_model,
  title={Liouvillian skin effect in an exactly solvable model},
  author={Yang, Fan and Jiang, Qing-Dong and Bergholtz, Emil J},
  journal={Physical Review Research},
  volume={4},
  number={2},
  pages={023160},
  year={2022},
  publisher={APS}
}

@article{NHSE_via_gainloss_in_opticlly_coupled_cavity_array,
title = {Non-Hermitian skin effect induced by on-site gain and loss in the optically coupled cavity array},
journal = {Results in Physics},
volume = {57},
pages = {107372},
year = {2024},
issn = {2211-3797},
doi = {https://doi.org/10.1016/j.rinp.2024.107372},
url = {https://www.sciencedirect.com/science/article/pii/S2211379724000548},
author = {Ming-Jie Liao and Mei-Song Wei and Zijian Lin and Jingping Xu and Yaping Yang}
}

@article{nonabelian_lattice_gaugefields_in_phtonic...,
  title={Non-Abelian lattice gauge fields in photonic synthetic frequency dimensions},
  author={Cheng, Dali and Wang, Kai and Roques-Carmes, Charles and Lustig, Eran and Long, Olivia Y and Wang, Heming and Fan, Shanhui},
  journal={Nature},
  volume={637},
  number={8044},
  pages={52--56},
  year={2025},
  publisher={Nature Publishing Group UK London}
}

@article{
synthesis&observation_of_nonAbelian_gauge_fields_in_real_space,
author = {Yi Yang  and Chao Peng  and Di Zhu  and Hrvoje Buljan  and John D. Joannopoulos  and Bo Zhen  and Marin Soljačić },
title = {Synthesis and observation of non-Abelian gauge fields in real space},
journal = {Science},
volume = {365},
number = {6457},
pages = {1021-1025},
year = {2019},
doi = {10.1126/science.aay3183},
URL = {https://www.science.org/doi/abs/10.1126/science.aay3183}}

@article{synthetic_non_abelian,
  title={Synthetic non-Abelian gauge fields for non-Hermitian systems},
  author={Pang, Zehai and Wong, Bengy Tsz Tsun and Hu, Jinbing and Yang, Yi},
  journal={Physical Review Letters},
  volume={132},
  number={4},
  pages={043804},
  year={2024},
  publisher={APS}
}

@article{twist_induced_NHSE_in_waveguides,
  title={Twist-induced non-Hermitian skin effect in optical waveguide arrays},
  author={Jiang, Chuang and Liu, Yang and Li, Xiaohong and Song, Yiling and Ke, Shaolin},
  journal={Applied Physics Letters},
  volume={123},
  number={15},
  year={2023},
  publisher={AIP Publishing}
}

@article{localisationtraninNHQM,
  title={Localization transitions in non-Hermitian quantum mechanics},
  author={Hatano, Naomichi and Nelson, David R},
  journal={Physical review letters},
  volume={77},
  number={3},
  pages={570},
  year={1996},
  publisher={APS}
}

@article{optical_force_enhancement,
  title={Optical force enhancement using an imaginary vector potential for photons},
  author={Descheemaeker, Lana and Ginis, Vincent and Viaene, Sophie and Tassin, Philippe},
  journal={Physical Review Letters},
  volume={119},
  number={13},
  pages={137402},
  year={2017},
  publisher={APS}
}

@article{critical_NHSE_Nature,
  title={Critical non-Hermitian skin effect},
  author={Li, Linhu and Lee, Ching Hua and Mu, Sen and Gong, Jiangbin},
  journal={Nature communications},
  volume={11},
  number={1},
  pages={5491},
  year={2020},
  publisher={Nature Publishing Group UK London}
}

@article{scaling_rule_for_cNHSE,
  title={Scaling rule for the critical non-Hermitian skin effect},
  author={Yokomizo, Kazuki and Murakami, Shuichi},
  journal={Physical Review B},
  volume={104},
  number={16},
  pages={165117},
  year={2021},
  publisher={APS}
}

@article{topological_funelling,
  title={Topological funneling of light},
  author={Weidemann, Sebastian and Kremer, Mark and Helbig, Tobias and Hofmann, Tobias and Stegmaier, Alexander and Greiter, Martin and Thomale, Ronny and Szameit, Alexander},
  journal={Science},
  volume={368},
  number={6488},
  pages={311--314},
  year={2020},
  publisher={American Association for the Advancement of Science}
}

@article{NHSE_in_UCatoms,
  title={Dynamic signatures of non-Hermitian skin effect and topology in ultracold atoms},
  author={Liang, Qian and Xie, Dizhou and Dong, Zhaoli and Li, Haowei and Li, Hang and Gadway, Bryce and Yi, Wei and Yan, Bo},
  journal={Physical review letters},
  volume={129},
  number={7},
  pages={070401},
  year={2022},
  publisher={APS}
}

@article{manipulate_NHSE_by_ORR,
  title={Manipulating the non-Hermitian skin effect in optical ring resonators},
  author={Xin, Haoran and Song, Wange and Wu, Shengjie and Lin, Zhiyuan and Zhu, Shining and Li, Tao},
  journal={Physical Review B},
  volume={107},
  number={16},
  pages={165401},
  year={2023},
  publisher={APS}
}

@article{engineering_NHSE_in_UCatoms,
  title={Engineering non-Hermitian skin effect with band topology in ultracold gases},
  author={Zhou, Lihong and Li, Haowei and Yi, Wei and Cui, Xiaoling},
  journal={Communications Physics},
  volume={5},
  number={1},
  pages={252},
  year={2022},
  publisher={Nature Publishing Group UK London}
}

@article{symmetry&topology,
  title={Symmetry and topology in non-Hermitian physics},
  author={Kawabata, Kohei and Shiozaki, Ken and Ueda, Masahito and Sato, Masatoshi},
  journal={Physical Review X},
  volume={9},
  number={4},
  pages={041015},
  year={2019},
  publisher={APS}
}

@article{NH_topology_ib_H_matter,
  title={Non-Hermitian topology in Hermitian topological matter},
  author={Hamanaka, Shu and Yoshida, Tsuneya and Kawabata, Kohei},
  journal={Physical Review Letters},
  volume={133},
  number={26},
  pages={266604},
  year={2024},
  publisher={APS}
}

@article{sanahal1,
  title={Gauge field induced skin effect in spinful non-Hermitian systems},
  author={Sanahal, Moirangthem and Panda, Subhasis and Nandy, Snehasish},
  journal={Physical Review B},
  volume={112},
  number={12},
  pages={125149},
  year={2025},
  publisher={APS}
}

@article{PW_Anderson,
  title={Absence of diffusion in certain random lattices},
  author={Anderson, Philip W and others},
  journal={Physical review},
  volume={109},
  number={5},
  pages={1492--1505},
  year={1958}
}

@article{AD_in_strongly_coupled,
  title={Anderson delocalization in strongly coupled disordered non-Hermitian chains},
  author={Jin, Wei-Wu and Liu, Jin and Wang, Xin and Zhang, Yu-Ran and Huang, Xueqin and Wei, Xiaomin and Ju, Wenbo and Yang, Zhongmin and Liu, Tao and Nori, Franco},
  journal={Physical Review Letters},
  volume={135},
  number={7},
  pages={076602},
  year={2025},
  publisher={APS}
}

@article{molignini,
  title={Anomalous skin effects in disordered systems with a single non-Hermitian impurity},
  author={Molignini, Paolo and Arandes, Oscar and Bergholtz, Emil J},
  journal={Physical Review Research},
  volume={5},
  number={3},
  pages={033058},
  year={2023},
  publisher={APS}
}

@article{NHSE_&_WN_in_disordered_NH_systems,
  title={Skin effect and winding number in disordered non-Hermitian systems},
  author={Claes, Jahan and Hughes, Taylor L},
  journal={Physical Review B},
  volume={103},
  number={14},
  pages={L140201},
  year={2021},
  publisher={APS}
}

@article{scaling_theory_of_localization,
  title = {Scaling Theory of Localization: Absence of Quantum Diffusion in Two Dimensions},
  author = {Abrahams, E. and Anderson, P. W. and Licciardello, D. C. and Ramakrishnan, T. V.},
  journal = {Phys. Rev. Lett.},
  volume = {42},
  issue = {10},
  pages = {673--676},
  numpages = {0},
  year = {1979},
  month = {Mar},
  publisher = {American Physical Society},
  doi = {10.1103/PhysRevLett.42.673},
  url = {https://link.aps.org/doi/10.1103/PhysRevLett.42.673}
}

@article{midya,
  title = {Topological phase transition in fluctuating imaginary gauge fields},
  author = {Midya, Bikashkali},
  journal = {Phys. Rev. A},
  volume = {109},
  issue = {6},
  pages = {L061502},
  numpages = {6},
  year = {2024},
  month = {Jun},
  publisher = {American Physical Society},
  doi = {10.1103/PhysRevA.109.L061502},
  url = {https://link.aps.org/doi/10.1103/PhysRevA.109.L061502}
}

@article{AL_transitions_in_disordered_NH,
  title={Anderson localization transitions in disordered non-Hermitian systems with exceptional points},
  author={Wang, C and Wang, XR},
  journal={Physical Review B},
  volume={107},
  number={2},
  pages={024202},
  year={2023},
  publisher={APS}
}

@article{AL_in_NH_AAH_model,
  title={Anderson localization in the non-Hermitian Aubry-Andr{\'e}-Harper model with physical gain and loss},
  author={Zeng, Qi-Bo and Chen, Shu and L{\"u}, Rong},
  journal={Physical Review A},
  volume={95},
  number={6},
  pages={062118},
  year={2017},
  publisher={APS}
}

@article{interplay_of_NHSE_and_AL,
  title={Interplay of non-Hermitian skin effects and Anderson localization in non-reciprocal quasiperiodic lattices},
  author={Jiang, Hui and Lang, Li-Jun and Yang, Chao and Zhu, Shi-Liang and Chen, Shu},
  journal={arXiv preprint arXiv:1901.09399},
  year={2019}
}

@article{Metal_Ins_phase_transition_in_NH_AAH_model,
  title={Metal-insulator phase transition in a non-Hermitian Aubry-Andre-Harper Model},
  author={Longhi, Stefano},
  journal={arXiv preprint arXiv:1908.03371},
  year={2019}
}

@article{Michael_Creutz,
  title = {End States, Ladder Compounds, and Domain-Wall Fermions},
  author = {Creutz, Michael},
  journal = {Phys. Rev. Lett.},
  volume = {83},
  issue = {13},
  pages = {2636--2639},
  numpages = {0},
  year = {1999},
  month = {Sep},
  publisher = {American Physical Society},
  doi = {10.1103/PhysRevLett.83.2636},
  url = {https://link.aps.org/doi/10.1103/PhysRevLett.83.2636}
}

@article{NH_multipole_SE_challenges_loc,
  title={Non-Hermitian multipole skin effects challenge localization},
  author={Gliozzi, Jacopo and Balducci, Federico and Hughes, Taylor L and De Tomasi, Giuseppe},
  journal={Physical Review B},
  volume={113},
  number={10},
  pages={L100203},
  year={2026},
  publisher={APS}
}

\newpage

\end{document}